\documentclass[aps,pra,twocolumn,groupedaddress,amsmath,amssymb]{revtex4-1}
\usepackage{graphicx}  % needed for figures
\usepackage{dcolumn}   % needed for some tables
\usepackage{bm}        % for math
\usepackage{verbatim}   % for math
\usepackage[usenames, dvipsnames]{color}
\usepackage{lipsum}
\usepackage{float}

\begin{document}

\title{ Maximal Steered Coherence Protection by Quantum Reservoir Engineering }

\author{Yusef Maleki}
\affiliation {Department of Physics and Astronomy, Texas A\&M University, 
College Station, Texas 77843-4242, USA}

\author{  Bahram Ahansaz}
\affiliation {
Physics Department, Azarbaijan Shahid Madani University,  Tabriz, Iran}

\date{\today}

\begin{abstract}
We show that the effects of decoherence on quantum steering ellipsoids can be controlled by a specific reservoir  manipulating, in both Markovian and non-Markovian realms. Therefore, the so-called maximal steered coherence could be protected through reservoir engineering implemented by coupling auxiliary qubits to the reservoir.
\end{abstract}

\pacs{}
\maketitle

\section{Introduction}
Quantum steering was first introduced by Schr\"{o}dinger in 1935 for analyzing the Einstein-Podolsky-Rosen (EPR) paradox \cite{Schrodinger1,Schrodinger2}.  This phenomenon, has attracted a remarkable renewed attention, due to its fundamental implications from a quantum information perspective and its significant role in quantum resources for information processors
 \cite{Wiseman,Saunders,Skrzypczyk}.

Quantum steering ellipsoids (QSEs), defined as the whole set of Bloch vectors to which Bob's qubit can be steered by all possible positive-operator valued measurements (POVMs) on Alice's qubit, was introduced by Jevtic, et. al., \cite{Jevtic} providing a faithful geometric representation of two-qubit states steering. In the same direction, the necessary and sufficient conditions for an ellipsoid to represent a two-qubit state were also derived \cite{Milne1}. Since QSE offers a useful visualization tool concerning quantum resources, it
has gained increased attention from quantum information viewpoints \cite{Jevtic, Milne1, Milne2, Shi1, Shi2, Nguyen1, Nguyen2, Quan, McCloskey}. The notion of QSE provides the tools to introduce the maximal steered coherence (MSC) as a measure determining the extent to which we can remotely create coherence via steering \cite{Hu}.
Considering the central importance of quantum coherence in various scenarios, spanning from the energy transport in biological systems \cite{Lloyd, Lambert} to quantum thermodynamics \cite{Aberg, Narasimhachar}, the connection between quantum steering and coherence reveals an important role of steering in quantum information processing. It is remarkable that, the interplay among coherence, which-path information and entanglement were recently investigated in the geometry of steriographic projection framework \cite{Maleki}.

Any realistic quantum system unavoidably interacts with its surrounding environment  which may enforce deleterious effects on the coherence of the system. Therefore, finding strategies to protect quantum coherence from unwanted interactions is a crucial task for the development of quantum-based technologies. In this regard, several strategies such as decoherence-free subspaces \cite{lid1, lid2}, quantum Zeno effect \cite{Facchi, Maniscalco} and weak measurement and quantum measurement reversal protocols have been proposed to control decoherence \cite{Basit,Huang}. These  strategies are relatively difficult as they mainly rely on operations on the main system. To circumvent this challenge,
alternative  schemes based on reservoir engineering has been developed. A remarkable example of such strategies could be adding some auxiliary subsystems into reservoir \cite{An, Ahansaz, Faizi}.

In this work, we show that  the dynamics of QSE could be protected by reservoir engineering leading to the protection of MSC through adding extra identical qubits to the reservoir. To demonstrate this, we consider a bipartite quantum system composed of two qubits possessed by Alice and Bob. We assume that  Alice's qubit does not interact with any external system, while Bob's qubit undergoes decoherence. We observe that MSC could be protected against decoherence by engineering reservoir through adding auxiliary qubits into the reservoir.

\section{QSE of a two-qubit state}
We start with a brief review of QSE and MSC. To this aim, we consider the general form of the two-qubit state $\rho_{AB}$, depicted in the Pauli basis
\begin{align}
    \rho_{AB}=\frac{1}{4} \bigg[I\otimes I+\mathbf{a}.\boldsymbol{\sigma}\otimes I+I\otimes \mathbf{b}.\boldsymbol{\sigma}+\sum_{m,n=1}^{3}T_{nm}\sigma_{n}\otimes \sigma_{m}\bigg],
\end{align}
where $I$ is the identity operator, and $\sigma_{j}$s, with $j=1,2,3$, are the three Pauli matrices. $\boldsymbol{\sigma}$ is the vector composed of these Pauli matrices.
Also, $\mathbf{a}=\mathrm{tr}(\rho_{AB} \boldsymbol{\sigma}\otimes I)$ and $\mathbf{b}=\mathrm{tr}(\rho_{AB} I\otimes \boldsymbol{\sigma})$ are the local Bloch vectors.
The bipartite correlations are determined by the matrix elements \cite{Horodecki}
\begin{eqnarray}
T_{nm}=\mathrm{tr}(\rho_{AB} \sigma_{n}\otimes \sigma_{m}).
\end{eqnarray}

If we perform a local measurement on Bob's qubit, Alice's state becomes steered. Hence, considering all possible local measurements by Bob, Alice's steering ellipsoid $\varepsilon_{A}$
is centered at \cite{Jevtic}
\begin{eqnarray}
C_{A}=\frac{\mathbf{a}-T\mathbf{b}}{1-\mathbf{b}^2}.
\end{eqnarray}
Thus, the ellipsoid matrix is given by
\begin{eqnarray}
Q_{A}=\frac{(T-\mathbf{a}\mathbf{b}^{T})}{1-\mathbf{b}^2}(1+\frac{\mathbf{b}\mathbf{b}^{T}}{1-\mathbf{b}^2})(T^{T}-\mathbf{b}\mathbf{a}^{T}),
\end{eqnarray}
where the eigenvalues of $Q_{A}$ are the squares of the ellipsoid semiaxes $s_{i}$ and its eigenvectors provide the orientation of these axes.

Alternatively, when Bob is steered by Alice's local measurements, we can obtain Bob's steering ellipsoid $\varepsilon_{B}$
by exchanging $A$ and $B$. Thus, his QSE ($\varepsilon_{B}$) is centred at
\begin{eqnarray}
C_{B}=\frac{\mathbf{b}-T^{T}\mathbf{a}}{1-\mathbf{a}^2},
\end{eqnarray}
and his ellipsoid matrix is
\begin{eqnarray}
Q_{B}=\frac{(T^{T}-\mathbf{b}\mathbf{a}^{T})}{1-\mathbf{a}^2}(1+\frac{\mathbf{a}\mathbf{a}^{T}}{1-\mathbf{a}^2})(T-\mathbf{a}\mathbf{b}^{T}).
\end{eqnarray}

Now, we investigate the set of so-called canonical states, which have particular importance in the steering ellipsoid formalism \cite{Jevtic, Milne1}.
This canonical state, $\widetilde{\rho}_{AB}$, corresponds to a two-qubit state in which Alice's marginal is maximally mixed.
We perform the stochastic local operations and classical communication (SLOCC) operator on qubit $A$ which transforms $\rho_{AB}$ to a canonical state
\begin{align}
\nonumber
   \rho_{AB}&\longrightarrow \widetilde{\rho}_{AB}
   =\big(\frac{1}{\sqrt{2\rho_{A}}}\otimes I \big)\rho\big(\frac{1}{\sqrt{2\rho_{A}}}\otimes I \big)^{\dag}
   \\
   &=\frac{1}{4} \bigg(I\otimes I+I\otimes \widetilde{\mathbf{b}}.\boldsymbol{\sigma}+\sum_{m,n=1}^{3}\widetilde{\mathbf{T}}_{nm}\sigma\otimes \sigma\bigg).
\end{align}

Since SLOCC operations on Alice do not affect Bob's steering ellipsoid, the same $\varepsilon_{B}$ can describe the characteristics of both $\rho_{AB}$ and the canonical state $\widetilde{\rho}_{AB}$.

Now, let us consider a bipartite quantum state $\rho_{AB}$, such that the eigenstates of  the reduced density matrix $\rho_{B}$ are denoted as $\Xi=\{|\chi_{i}\rangle\}$.
When we perform a positive operator-valued measure (POVM) on Alice and obtain an outcome $M$, Bob's state is steered to $\rho^{M}_{B}:=\mathrm{tr}_{A}(M\otimes I \rho)/p_{M}$ with $p_{M}:=\mathrm{tr}(M\otimes I \rho)$, where $I$ represents the single qubit identity operator. The quantum coherence of the steered states $\rho^{M}_{B}$, as the summation of the absolute values of off-diagonal elements in the basis $\{|\chi_{i}\rangle\}$,  gives \cite{Baumgratz}
\begin{eqnarray}
C(\rho^{M}_{B},|\chi_{i}\rangle)=\frac{1}{p_{M}}\sum_{i \neq j}|\langle \chi_{i}|\mathrm{tr}_{A}(M\otimes I \rho)|\chi_{j}\rangle|.
\end{eqnarray}
By considering all possible POVM operators on Alice, the set of all $\rho^{M}_{B}$ provides $\varepsilon_{B}$.
Maximizing the coherence $C(\rho^{M}_{B},|\chi_{i}\rangle)$ over all possible POVM operators $M$ gives MSC as \cite{Hu}
\begin{align}
  MSC(\rho^{M}_{B}):=\max_{M \in POVM} \left[\frac{1}{p_{M}}\sum_{i \neq j}|\langle \chi_{i}|\mathrm{tr}_{A}(M\otimes I \rho)|\chi_{j}\rangle|\right].
\end{align}

If $\rho^{M}_{B}$ is degenerate, $\Xi$ is not uniquely defined; however, MSC is defined over all possible POVM operators and taking  infimum over all possible reference basis $\Xi$ as \cite{Hu}
\begin{align}
\nonumber
&MSC(\rho^{M}_{B}):=
\\
&\inf_{\Xi} \left\{ \max_{M \in POVM} \left[\frac{1}{p_{M}}\sum_{i \neq j}|\langle \chi_{i}|\mathrm{tr}_{A}(M\otimes I \rho)|\chi_{j}\rangle|\right] \right\}.
\end{align}

According to Ref. \cite{Hu}, MSC is the maximal perpendicular distance between a point on the surface of $\varepsilon_{B}$ and the reference basis.
Specifically, when the input state is an X-state and reference basis lies along an axis of $\varepsilon_{B}$, it was found that MSC is the length of the longest of the other
two semiaxes. However, when the reference basis does not lies along the axis of $\varepsilon_{B}$, the MSC can be simply expressed by the length of the longest semiaxes of Bob's steering ellipsoid.

\section{Dynamics of $N$ two-level systems in a common reservoir}

We investigate a system of $N$ identical qubits (two-level atoms) immersed in a common zero-temperature reservoir that have no direct interactions with each other. The Hamiltonian of the whole system can be written as ($\hbar=1$)
\begin{align}
\nonumber
   H=&\sum_{j=1}^{N} {\Omega_{0} \sigma_{j}^{+} \sigma_{j}^{-}}+\sum_{k} \omega_{k} b_{k}^{\dagger} b_{k}
   \\
+& \sum_{j=1}^{N} \sum_{k} \big(g_{k}^{j}{\sigma_{j}^{+}} b_{k}+{g_{k}^{j}}^{*}{\sigma_{j}^{-}} b_{k}^{\dagger}\big).
\end{align}
Here, $\sigma_{j}^{-}=(\sigma_{j}^{+})^{\dagger} \equiv |0\rangle \langle j|$ ($j=1,2,...,N$) is the lowering operator, having the transition frequency $\Omega_{0}$ between the excited state of $j$th atom and the ground state $|0\rangle \equiv |0\rangle^{\otimes N}$, where $|j\rangle\equiv|000...1_{j}...0\rangle$
is the typical standard basis state for the $N$-fold single excitation subspace.
Also, $a_{k}$ ($a_{k}^\dag$) is the annihilation (creation) operator of the $k$th field mode with frequency $\omega_{k}$. $g_{k}^{j}$ is the coupling strength between the $k$th field mode and the $j$th atom.

Now, let us assume $g_{k}^{j}=\frac{1}{\sqrt{N}} G_{k}$, for $j=1,2,...,N$ and define alternative set of basis as
\begin{eqnarray}
|\varphi_{l}\rangle=\frac{1}{\sqrt{N}} \sum_{m=0}^{N-1} \mathrm{exp}(\frac{2i\pi ml}{N}) |m+1\rangle,
\end{eqnarray}
for $l=0,1,...,N-1$.
Then, we can express the total Hamiltonian in Eq. (1) as
\begin{align}
\nonumber
   \mathcal{H}=\sum_{j=0}^{N-1} {\Omega_{0} |\varphi_{j}\rangle \langle\varphi_{j}|}+\sum_{k} \omega_{k} b_{k}^{\dagger} b_{k}
   \\
    +\sum_{k} \big(G_{k}|\varphi_{0}\rangle \langle0|b_{k}+G_{k}^{*} |0\rangle \langle\varphi_{0}|b_{k}^{\dagger} \big).
    \label{Hamiltonian2}
\end{align}
This Hamiltonian suggests that the interaction between the system and the reservoir is restricted only to the two-dimensional subspace of the eigenstate $|\varphi_{0}\rangle$ and the ground state, i.e., $\{|0\rangle, |\varphi_{0}\rangle \}$.
Therefore, the other $N-1$ eigenstates $\{|\varphi_{1}\rangle, |\varphi_{2}\rangle, ..., |\varphi_{N-1}\rangle \}$ remain uncoupled to the reservoir, forming a decoherence-free subspace.
In other words, the total Hamiltonian given by Eq. \eqref{Hamiltonian2}, consists of the noisy subspace, corresponding to the interaction of the qubits with the reservoir, and the $(N-1)$-dimensional free subspace.

Having Eq. \eqref{Hamiltonian2} in hand, we consider the dynamics of a system of $N$ two-level atoms,  where  the system is initially (at time $t=0$) described by the state
\begin{eqnarray}
  |\psi(0)\rangle=c_{0}(0)|0\rangle_{S} |0\rangle_{E}+\sum_{j=0}^{N-1}C_{j}(0)|\varphi_{j}\rangle_{S} |0\rangle_{E}.
\end{eqnarray}
Since the Hamiltonian conserves the number of excitations in the whole system, the time-evolved state $|\psi(t)\rangle$ can be found as
\begin{align}
\nonumber
    |\psi(t)\rangle=& c_{0}(t)|0\rangle_{S} |0\rangle_{E}+\sum_{j=0}^{N-1}C_{j}(t)|\varphi_{j}\rangle_{S} |0\rangle_{E}
    \\
    &+\sum_{k} D_{k}(t) |0\rangle_{S} |1_{k}\rangle_{E},
\end{align}
where $|1_{k}\rangle_{E}$ denotes the state of the reservoir with only one excitation in the $k$th mode. By using Schr\"{o}dinger equation
\begin{equation}
i\frac{d}{dt}|\psi(t)\rangle=\mathcal{H}|\psi(t)\rangle,
\end{equation}
the time-dependent coefficients $\widetilde{C}_{j}(t)$ and $\widetilde{D}_{k}(t)$ are determined through
\begin{equation}
\begin{array}{l}
\displaystyle \frac{d C_{0}(t)}{d t}=-i\Omega_{0}C_{0}(t)-i\sum_{k}G_{k}D_{k}(t),\\\\
\displaystyle \quad \frac{d D_{k}(t)}{d t}=-i\omega_{k}D_{k}(t)-i G_{k}^{*}C_{0}(t).
\end{array}
\label{eq:xdef}
\end{equation}
And the remaining coefficients are governed by $\frac{d C_{j}(t)}{d t}=-i\Omega_{0}C_{j}$, for $j=1,2,...,N-1$.
Also, it should be noted that the fact $\mathcal{H}|0\rangle_{S} |0\rangle_{E}=0$, dictates $c_{0}(t)=c_{0}(0)$. Also,
introducing the new set of coefficients $\tilde{C}_{j}(t)=e^{i\Omega_{0}t}C_{j}(t)$ and $\tilde{D}_{k}(t)=e^{i\omega_{k}t}D_{k}(t)$ results in
\begin{equation}
\begin{array}{l}
\displaystyle \frac{d \tilde{C}_{0}(t)}{d t}=-i\sum_{k}G_{k}e^{i(\Omega_{0}-\omega_{k})t}\tilde{D}_{k}(t),\\\\
\displaystyle \quad \frac{d \tilde{D}_{k}(t)}{d t}=-iG_{k}^{*}e^{-i(\Omega_{0}-\omega_{k})t}\tilde{C}_{0}(t),
\end{array}
\label{eq:xdef}
\end{equation}
where $\tilde{C}_{j}(t)=\tilde{C}_{j}(0)$, for $j=1,2,...,N-1$. Thus, these coefficients are time-independent.
Integrating the second relation of Eq. (18) and substituting it into the first one provides a closed form integro-differential equation for $\tilde{C}_{0}(t)$ as
\begin{eqnarray}
\frac{d \tilde{C}_{0}(t)}{d t}=-\int_{0}^{t} f(t-t')\tilde{C}_{0}(t')dt'.
\end{eqnarray}
In this equation, the correlation function $f(t-t') $ can be expressed as
\begin{eqnarray}
f(t-t')=\int_{-\infty}^{\infty}d\omega \mathfrak{J}(\omega)e^{i(\Omega_{0}-\omega)(t-t')},
\end{eqnarray}
where
\begin{align}
\nonumber
  \mathfrak{J}(\omega)=&\sum_{k}|G_{k}|^2\delta(\omega-\omega_{k})
  \\
    =& N\sum_{k}|g_{k}|^2\delta(\omega-\omega_{k})=NJ(\omega),
\end{align}
is the spectral density of reservoir. Eq. (19) can be solved using Laplace transformation after specifying the spectral density.

In our analyses, we take the Lorentzian spectral distribution with the structure function as
\begin{eqnarray}
J(\omega)=\frac{1}{2\pi}\frac{\gamma_{0}\lambda}{(\omega-\Omega_{0})^2+\lambda^2},
\end{eqnarray}
where $\lambda$ is the spectral width, $\gamma_{0}$ is the coupling strength, and $\Omega_{0}$ is the central frequency of the reservoir.
Using Laplace transformation and its inverse, the exact solution of the probability amplitudes ${C}_{0}(t)$ can be obtained as
\begin{eqnarray}
\tilde{C}_{0}(t)=e^{-\lambda t/2}\Big(\mathrm{cosh}(\frac{Dt}{2})+\frac{\lambda}{D}\mathrm{sinh}(\frac{Dt}{2})\Big)\tilde{C}_{0}(0),
\end{eqnarray}
where $D=\sqrt{\lambda^{2}-2N\gamma_{0}\lambda}$. Now, we return to the old basis $\{|j\rangle\}$ and find $|\psi(t)\rangle$ in Eq. (15)  as
\begin{align}
\nonumber
  |\psi(t)\rangle=&c_{0}(t)|0\rangle_{S} |0\rangle_{E}+\sum_{j=0}^{N-1}\chi_{j}(t)|j+1\rangle_{S} |0\rangle_{E}
  \\
  +&\sum_{k} D_{k}(t) |0\rangle_{S} |1_{k}\rangle_{E},
\end{align}

where
\begin{eqnarray}
\chi_{j}(t)=\frac{1}{\sqrt{N}}\sum_{m=0}^{N-1} \mathrm{exp}(\frac{2i\pi m j}{N}) C_{m}(t)
\end{eqnarray}
is the probability amplitude for the excitation of the ($j+1$)th atom.

\section{QSE and MSC Protection}

\begin{figure}
\centering
\includegraphics[width=8cm]{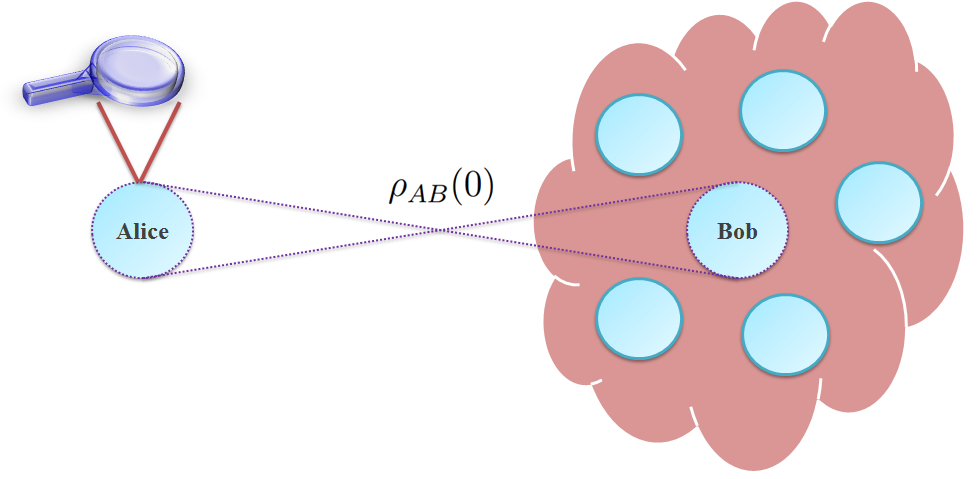}
\caption{Schematic  of the system: the bipartite state $\rho_{AB}(0)$ is shared by Alice and Bob. Bob's party undergoes decoherence by interacting with its local environment.
 Once performing a local measurement by Alice, Bob's state gets steered. Adding auxiliary qubits protects Bob's MCS.}
\label{fig1}
\end{figure}

 Now, we are going to demonstrate the effects of decoherence on QSE and MSE, and show how auxiliary qubits can aid in system protection. To this end, we assume two qubits (two-level atoms) belonging to Alice (A) and Bob (B), where Alice's atom has no direct interaction with the reservoir but Bob's atom interacts with its local environment (see Fig. \ref{fig1}). We further allow some auxiliary atoms to interact with this common reservoir, as considered in  the previous section.
Accordingly, we assume the first atom ($j$=1) to be Bob's system and the remaining atoms to be the auxiliary qubits. We assume all the auxiliary atoms to be initially in their ground states, and only Bob's qubit could be excited. Then, by considering the initial probability amplitudes $\chi_{1}(0)=\chi_{2}(0)=\chi_{N-1}(0)=0$, the time-evolved probability amplitude $\chi_{0}(t)$ can be derived as $\chi_{0}(t)=\mathcal{G}(t) \chi_{0}(0)$, where
\begin{align}
   \mathcal{G}(t)&=e^{-i\Omega_{0}t} \left[\frac{N-1}{N}+\frac{e^{-\lambda t/2}}{N} \Big(\mathrm{cosh}{(\frac{Dt}{2})}+\frac{\lambda}{D} \mathrm{sinh}{(\frac{Dt}{2})}\Big)\right].
\end{align}
One can express the dynamics of the Bob's system in the operator-sum representation as
\begin{align}
\rho_{Bob}(t)=\sum_{i=0}^{1} \mathcal{K}_{i}(t) \rho_{Bob}(0) \mathcal{K}_{i}(t)^{\dag},
\end{align}
where the corresponding Kraus operators $\mathcal{K}_{i}(t)$ can be written as
\begin{align}
\mathcal{K}_{0}(t)=\left(
                \begin{array}{cc}
                  1 & 0\\\\
                  0 & \sqrt{p(t)}\\
                \end{array}
              \right),\quad \mathcal{K}_{1}(t)=\left(
                \begin{array}{cc}
                  0 &  \sqrt{1-p(t)}\\\\
                  0 & 0\\
                \end{array}
              \right),
\end{align}
with $\sum_{i=0}^{1} \mathcal{K}_{i}^{\dag}(t) \mathcal{K}_{i}(t)=I$, for all values of $t$. For which we have $p(t)=|\mathcal{G}(t)|^2$.

Now,  we assume that Alice and Bob share the initial state of the form
\begin{eqnarray}
\rho_{AB}(0)=q |\varphi\rangle \langle\varphi|+\frac{1-q}{4}I\otimes I,
\end{eqnarray}
with
\begin{eqnarray}
|\varphi\rangle=\cos(\frac{\theta}{2})|10\rangle+\sin(\frac{\theta}{2})|01\rangle,
\end{eqnarray}
where, $0\leq q\leq 1$. We note that this is a somewhat general setting, where the initial state could be chosen to be a pure or a mixed state. Hence, the dynamics of the bipartite shared state $\rho_{AB}(t)$ can be obtained as
\begin{eqnarray}
\rho_{AB}(t)=\sum_{i=0}^{1} (I\otimes \mathcal{K}_{i}(t))\rho_{AB}(0) (I\otimes \mathcal{K}_{i}(t)^{\dag}),
\end{eqnarray}

Once we identified the density matrix $\rho_{AB}(t)$, we can proceed with investigating the steering of one of the parties.  According to section II,  if Alice performs a measurement on her qubit, Bob's QSE shall be centered at

\begin{align}
\nonumber
&C_{B}=
\\
&\big(0, 0, \displaystyle \frac{\cos(\theta)[p-\cos(\theta)+p\cos(\theta)]q^2-p\cos(\theta)q-p+1}{1-q^2 \cos^2(\theta)}\big),
\end{align}
and ellipsoid matrix of Bob shall read
\begin{align}
\nonumber
&Q_{B}=
\\
&\left(
                \begin{array}{ccc}
                  \displaystyle \frac{q^2 p \sin^2(\theta)}{1-q^2 \cos^2(\theta)} & 0 & 0\\
                  0 & \displaystyle \frac{q^2 p \sin^2(\theta)}{1-q^2 \cos^2(\theta)} & 0\\
                  0 & 0 & \displaystyle \frac{q^2 p^2 (1-q\cos^2(\theta))^2}{(1-q^2 \cos^2(\theta))^2}\\
                \end{array}
              \right).
\end{align}

Therefore, the lengths of the semiaxes in $x_{1}$, $x_{2}$ and $x_{3}$  directions, denoted respectively as $s_1$, $s_2$ and $s_3$, are
\begin{equation}
\begin{array}{l}
\displaystyle s_{1}=s_{2}=\frac{q \sqrt{p}\sin(\theta)}{\sqrt{1-q^2 \cos^2(\theta)}},\\\\
\displaystyle \quad s_{3}=\frac{q p (1-q\cos^2(\theta))}{1-q^2 \cos(\theta)^2},
\end{array}
\end{equation}
and QSE of Bob, steered by Alice, reads
\begin{eqnarray}
\varepsilon_{B}=\left\{\left(
                \begin{array}{ccc}
                  C_{B}(1)\\
                  C_{B}(2)\\
                  C_{B}(3)\\
                \end{array}
              \right)+\left[
                \begin{array}{ccc}
                  s_{1} x_{1}\\
                  s_{2} x_{2}\\
                  s_{3} x_{3}\\
                \end{array}
              \right] |x\leq1 \right\}.
\end{eqnarray}

As we addressed earlier, the length of the longest semiaxes of Bob's steered state $\rho^{M}_{B}$ can represent MSC, hence, providing
\begin{eqnarray}
MSC(\rho^{M}_{B})=\max\{s_{1}, s_{2}, s_{3}\}.
\end{eqnarray}

\begin{figure*}
\includegraphics[width=\linewidth]{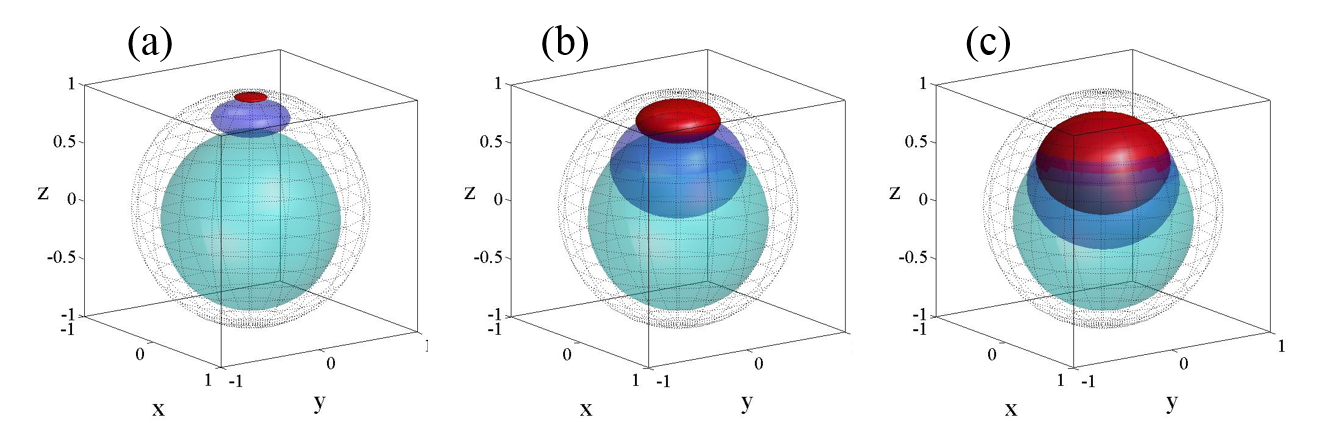}
\caption{
The dynamics of Bob's quantum steering ellipsoid, $\varepsilon_{B}$, for $q=0.8$ and $\theta=\pi/3$. (a) $N=1$, (b) $N=3$ and (c) $N=6$.
The green ellipsoids depicts the initial $\varepsilon_{B}$, and the red and the blue ellipsoids represent $\varepsilon_{B}$ at $t=0.8$ an $t=1.6$, respectively.
The parameters of the reservoir are fixed as $\lambda=1$, $\gamma_{0}=8$.}
\label{fig:2}
\end{figure*}

\begin{figure*}
\includegraphics[width=\linewidth]{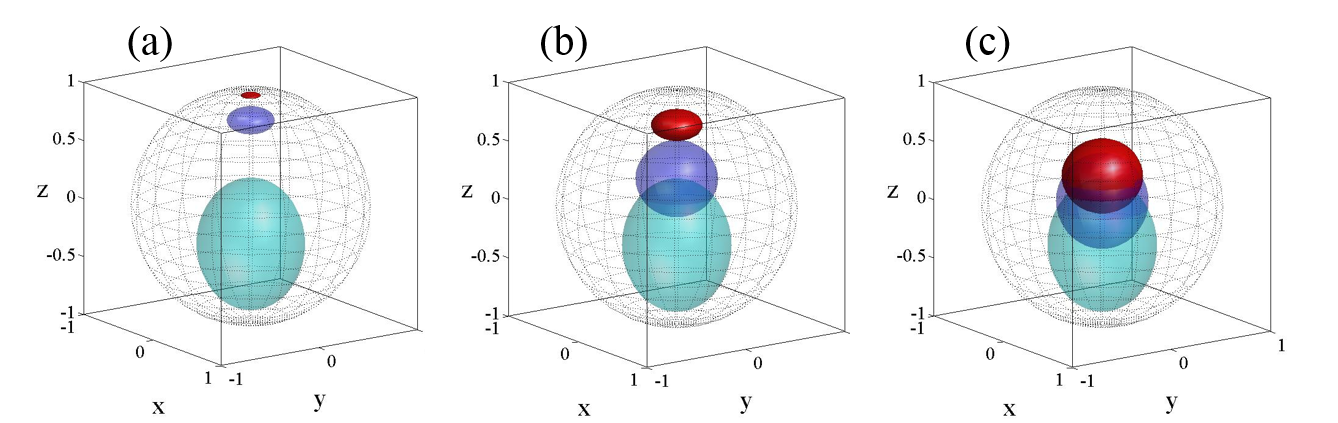}
\caption{
The dynamics of Bob's quantum steering ellipsoid, $\varepsilon_{B}$, for $q=0.8$ and $\theta=\pi/8$. (a) $N=1$, (b) $N=3$ and (c) $N=6$.
The green ellipsoids depicts the initial $\varepsilon_{B}$, and the red and the blue ellipsoids represent $\varepsilon_{B}$ at $t=0.8$ an $t=1.6$, respectively.
The parameters of the reservoir are fixed as $\lambda=1$, $\gamma_{0}=8$.}
\label{fig:3}
\end{figure*}

With these analyses, we present the dynamics of $\varepsilon_{B}$ in Figs. \ref{fig:2} \&  \ref{fig:3}, for $\theta=\pi/3$ and $\theta=\pi/8$, respectively. In both of these figures,  green, red and the blue ellipsoid represents $\varepsilon_{B}$ at $t=0, $ $ 0.8$ and $1.6$, respectively. Comparing  $t=0$ and $t=0.8$ ellipsoids (the green and the red ones), the semiaxes lengths reduces and the ellipsoids shrink due to decoherence and relaxation of the system. However, the ellipsoids at $t=1.6$ are larger compared to the ones at $t=0.8$. This is due to the information back flow from the reservoir to the system. According to Figs. \ref{fig:2} \&  \ref{fig:3}, adding auxiliary qubits increases semiaxes lengths of the  ellipsoids, thus, enhancing protection of the system.  This protection improves by increasing the number of auxiliary qubits and a full preservation of the system can be attained at the limit of very large $N$ (this is evident from Eq. (26)). Furthermore, from Eq. (34) and comparing Figs. \ref{fig:2} and  \ref{fig:3}, we find the the steered state is highly dependant on $\theta$.  To further analyse the dependence of the ellipsoid to $\theta$, let us focus on the region $0 \leq \theta \leq \pi/2$, where the state $|\varphi\rangle$ in Eq. (30) changes from the separable state to the maximally entangled state. In this region, the semiaxes lengths given by Eq. (34) are monotonically increasing functions in terms of $\theta$. Therefore, lengths of the semiaxes  are minimum for $\theta=0$ (which are given by $s_1=s_2=0$ and $s_3=qp/(1+q)$), and attain their maximum for $\theta=\pi/2$ (which are given by $s_1=s_2=q\sqrt{p}$ and $s_3=qp$). Note that, for $q=1$ at $t=0$, the steered state ellipsoid reduces to Bloch sphere, as expected. Thus, increasing $0 \leq \theta \leq \pi/2$ enlarges the ellipsoid, which could be visualized  by comparing Figs. \ref{fig:2} and  \ref{fig:3}.

\begin{figure}
\centering
\includegraphics[width=7 cm]{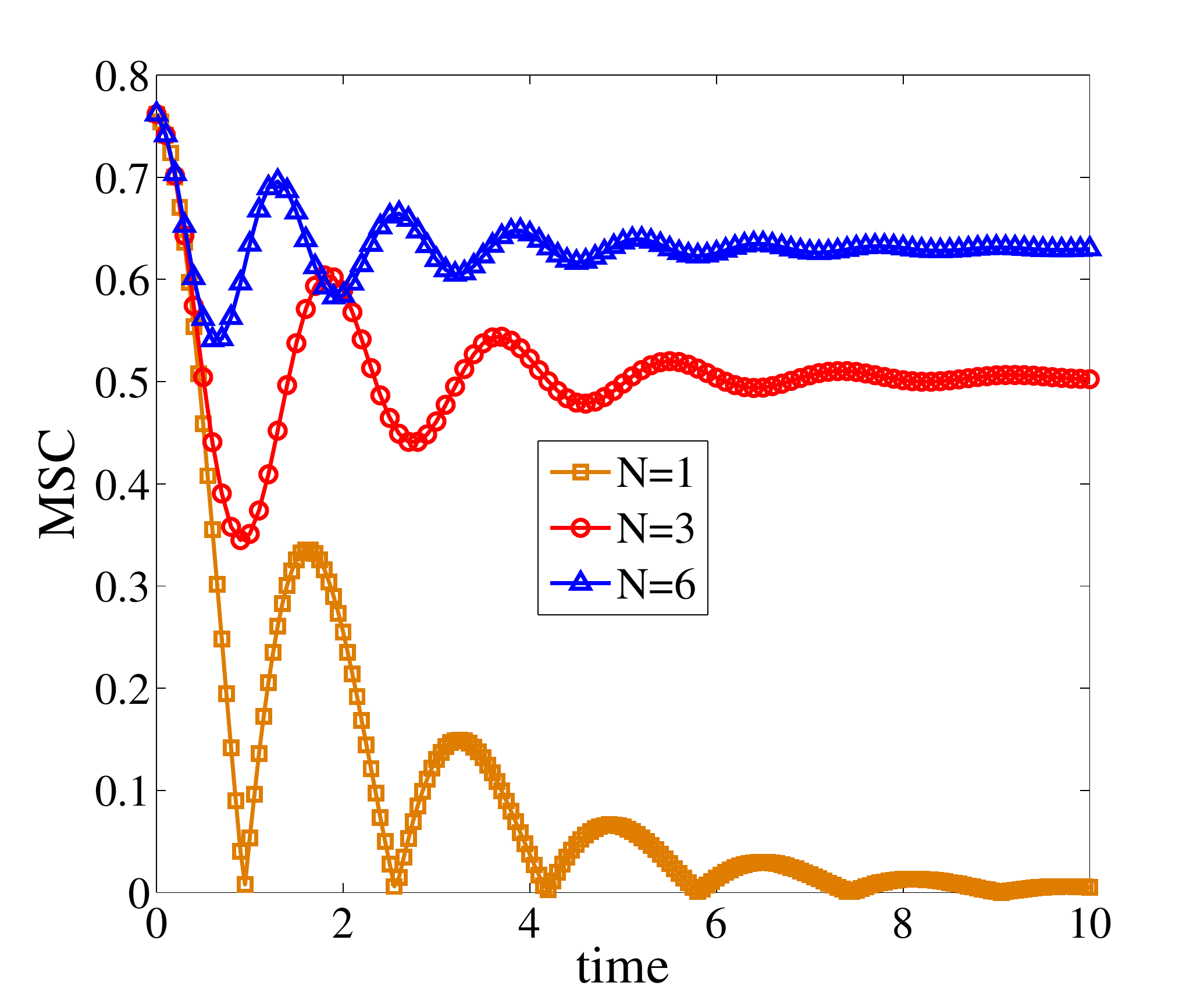}
\caption{ The dynamics of MSC versus time with the initial conditions $q=0.8$ and $\theta=\pi/3$,  for $N=1, 3$ and 6.
The parameters of the reservoir are fixed as $\lambda=1$, $\gamma_{0}=8$.}
\end{figure}

\begin{figure}
\centering
\includegraphics[width=7 cm]{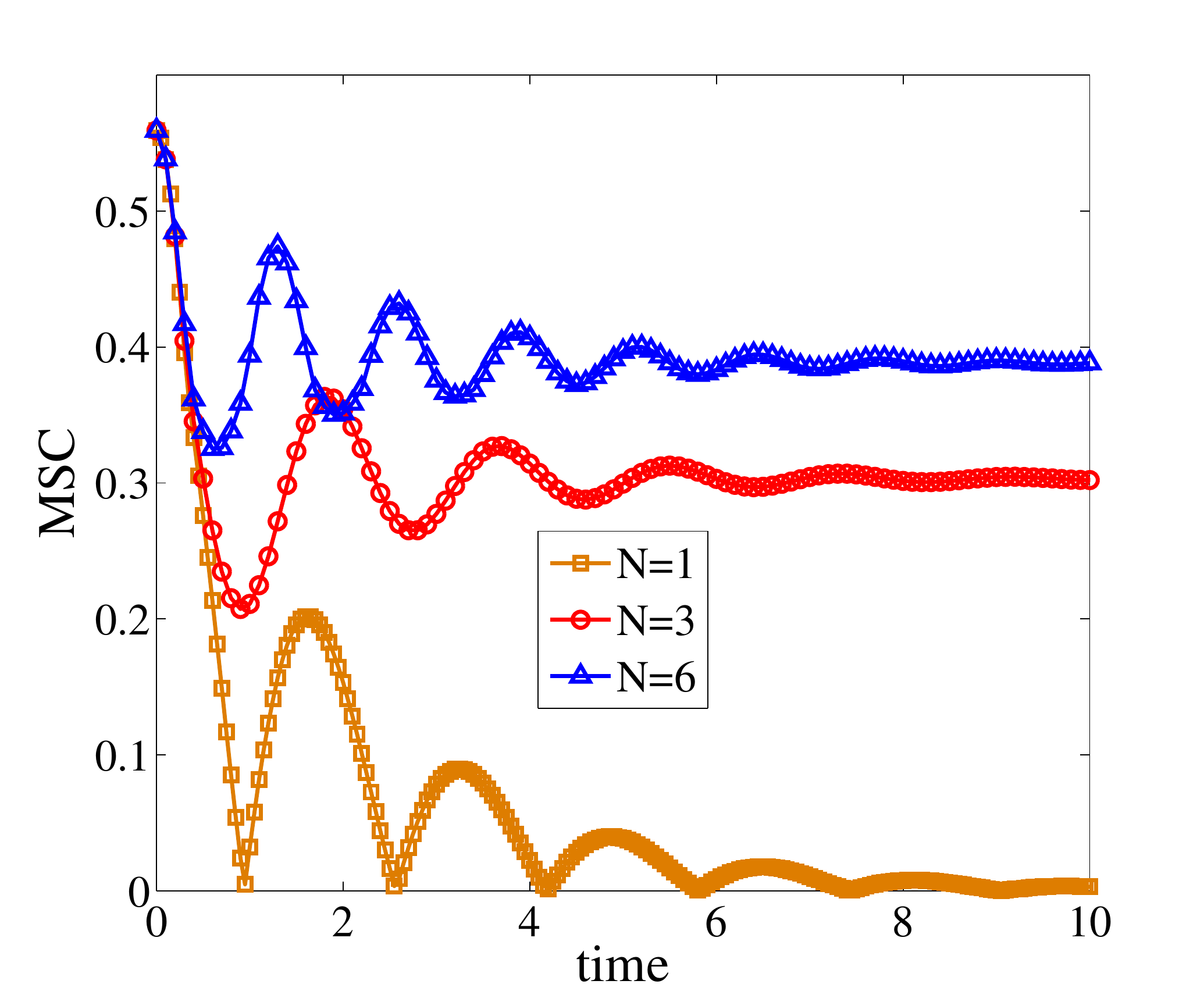}
\caption{ The dynamics of MSC versus time with the initial conditions $q=0.8$ and $\theta=\pi/8$,  for $N=1, 3$ and 6.
The parameters of the reservoir are fixed as $\lambda=1$, $\gamma_{0}=8$.}
\label{fig4}
\end{figure}

The  dynamics of MSC corresponding to the parameters specified in Figs. 2 \& 3 are depicted in Figs. 4 \& 5, respectively. Supporting our analyses above, the oscillatory dynamics of these figures show the flow and back flow of information between the system and the reservoir. Also, in the absence of auxiliary atoms ($N=1$), the ellipsoids disappear in the north pole when $t$ is large enough (see Fig. 2(a) and 3(a)).
Hence,  as depicted in Fig. 4 \& 5, MSC approaches zero after some oscillatory dynamics. However, adding the auxiliary qubits into the reservoir
inflates $\varepsilon_{B}$, enabling protection of the system  (see Figs. 2-5). Therefore, as we observe from Figs. 4 \& 5, the specific engineering of the reservoir through auxiliary qubits can, in fact, protect MSC from decoherence.

To understand how auxiliary qubits assist protecting QSE and MSC, we briefly elaborate on the connection between the proposed scheme of this work and the notion of decoherence-free subspace \cite{lid1, lid2}. As was shown earlier, the initial state of the system can be considered as $|\psi(0)\rangle_{S}=c_{0}(0)|0\rangle_{S}+C_{0}(0)|\varphi_{0}\rangle_{S}$ for $N=1$. However, for $N=3$ and $N=6$ it could be expressed as  $|\psi(0)\rangle_{S}=c_{0}(0)|0\rangle_{S}+C_{0}(0)|\varphi_{0}\rangle_{S}+C_{1}(0)|\varphi_{1}\rangle_{S}+C_{2}(0)|\varphi_{2}\rangle_{S}$ and
$|\psi(0)\rangle_{S}=c_{0}(0)|0\rangle_{S}+C_{0}(0)|\varphi_{0}\rangle_{S}+...+C_{5}(0)|\varphi_{5}\rangle_{S}$, respectively. As was mentioned in section III, the interaction of the system with the reservoir is only restricted to the ground state $|0\rangle_{S}$ and the exited state $|\varphi_{0}\rangle_{S}$, that leaves the bases $\{|\varphi_{1}\rangle_{S}, |\varphi_{2}\rangle_{S}\}$ for $N=3$ and $\{|\varphi_{1}\rangle_{S}, ..., |\varphi_{5}\rangle_{S}\}$ for $N=6$ uncoupled, which form a decoherence-free subspace. However,
the initial state $|\psi(0)\rangle_{S}$ has a support on the these uncoupled bases, and since the dimension of the decoherence-free subspace becomes larger by increasing the number of the auxiliary atoms, the process
protects MSC from decoherence. When the number of auxiliary atoms is very large, the decoherence-free subspace becomes sufficiently large such that the initial state of the respective qubit  belongs completely to this subspace and remains unchanged.

%\vspace{1cm}
\section{Conclusion}

In this paper, we demonstrated that maximal steered coherence could be protected by reservoir engineering, enabling preservation of quantum steering ellipsoid. Our reservoir was considered to be at zero temperature, having Lorentzian spectral distribution. More specifically, the MSC could protected in both Markovian and non-Markovian scenarios, offering a powerful technique for suppressing decoherence for information processing application.

\end{document}